\documentclass{pasj02}
\usepackage{url}
\usepackage{diagbox}

\jyear{2025}
\Received{}
\Accepted{}

\begin{document} 

\title{ Huntsman may need to be irradiated }

\author{
 Shunyi \textsc{Lan},\altaffilmark{1,2}\altemailmark\orcid{0009-0001-8816-8523} \email{lanshunyi@ynao.ac.cn} 
 Xiangcun \textsc{Meng},\altaffilmark{1}\altemailmark \orcid{0000-0001-5316-2298} \email{xiangcunmeng@ynao.ac.cn} 
}
\altaffiltext{1}{International Centre of Supernovae (ICESUN), Yunnan Key Laboratory of Supernova Research, Yunnan Observatories, Chinese Academy of Sciences (CAS), Kunming 650216, China}
\altaffiltext{2}{University of Chinese Academy of Sciences, Beijing 100049, PR China}


\maketitle

\begin{abstract}
Millisecond pulsars are rapidly rotating neutron stars, and it is now widely accepted that their extremely short rotation periods result from the accretion of material from a companion star. Binary evolution theory predicts that millisecond pulsars can have various types of companion stars. However, in observations, binary pulsars with giant companions, referred to as ``huntsman pulsars'', are extremely rare. Following the initial discovery of the first huntsman pulsar, 1FGL J1417.7-4407, a second huntsman millisecond pulsar binary, PSR J1947-1120, has been recently reported approximately a decade later. In this paper, we model the formation and evolution of two huntsman pulsars. Our model with the irradiation effect can explain the observed properties of huntsman pulsar binaries and suggests that if the irradiation effect is considered, the companion star may be a normal red giant star, rather than just a red bump star.
\end{abstract}


\section{Introduction} \label{sec1}

Millisecond pulsars (MSPs) are rapidly rotating neutron stars (NSs). In the recycling scenario, an NS in a binary system can accrete material to spin itself up. At the same time, the NS's magnetic field decays \citep{1989Natur.342..656S,1992ASIC..377.....V,2023pbse.book.....T}. This theory appears natural, as most MSPs are found in binary systems. Moreover, the discoveries of transitional MSPs and accreting millisecond X-ray pulsars provide strong support for the recycling scenario \citep{2009Sci...324.1411A,2013Natur.501..517P,2014MNRAS.441.1825B,2021ASSL..461..143P}.

In principle, the types of companion stars in binary MSPs can vary significantly: main-sequence stars, giant stars, helium stars, white dwarfs, NSs, and black holes. Many of these have been observed, except for helium stars and black holes as companions to MSPs. Binary MSPs with giant companions are referred to as ``huntsman'' pulsars, which are extremely rare. \cite{2015ApJ...804L..12S} reported the first huntsman pulsar 1FGL J1417.7–4407 (hereafter J1417). J1417 is in a 5.4 day orbit with a giant companion star of mass $0.28_{-0.03}^{+0.07}~M_{\odot}$ and has a spin period 2.66 ms \citep{2015ApJ...804L..12S,2016ApJ...820....6C,2025arXiv250105509S}. {J1417 exhibits some characteristics similar to redback pulsars; however, its orbital period of 5.4 days is significantly longer than the typical 0.1-1 day range observed in classical redback systems \citep{roberts2012surrounded}.} Therefore, J1417 and similar binary pulsars with giant companions were named ``huntsman'' after another type of spider and are considered a new subclass of MSPs \citep{2017ApJ...851...31S}. \cite{2017ApJ...851...31S} reported a NS+giant star binary, 2FGL J0846.0+2820, but no MSP was detected in this system, leaving it as a huntsman candidate. Recently, \cite{2025arXiv250105509S} reported the second confirmed huntsman pulsar, PSR J1947–1120 (hereafter J1947). J1947 has a spin period of 2.24 ms and is in a 10.265 day orbit with a red giant companion of mass $0.32\pm0.03~M_\odot$.

In the report by \cite{2025arXiv250105509S}, the companion star of J1947 could only be explained as a red bump star based on binary evolution modeling. A red-bump star refers to a red giant observed during the red giant branch bump phase, when its hydrogen-burning shell encounters the chemical discontinuity created by the first dredge-up \citep{2013sse..book.....K}. This interaction temporarily slows or slightly reverses the star's luminosity increase, producing a localized overdensity (the ``bump'') in the luminosity function. If a red giant in a mass-transferring binary acts as the donor star, the mass transfer will be temporarily halted when the star enters the red-bump phase. This occurs because the star's radius decreases during this phase, causing it to detach from its Roche lobe \citep{2024A&A...690A..88L}. The distinctive characteristics of huntsman pulsars lie in the absence of ongoing mass transfer and the presence of a red giant companion (in contrast to low-mass X-ray binaries where active mass transfer is occurring). A key requirement for explaining the formation of huntsman pulsars is the existence of a mechanism that causes the companion star to detach from its Roche lobe, thereby halting the mass transfer process. While previous studies suggested that the companion star of J1947 must be a red bump star, our work shows that a normal red giant star is also a viable explanation when irradiation effects are considered.

The irradiation effect may occurs when energies released from an accreting compact objects heat the companion star’s outer layers, disrupting its thermal equilibrium and inducing cyclic mass transfer \citep{1991Natur.350..136P,2012ApJ...753L..33B,ginzburg2021novae}. This process shortens the duration of the low-mass X-ray binary phase and modifies binary evolution pathways. Crucially, the irradiation effect offers a potential resolution to the long-standing birth rate discrepancy between MSPs and their progenitor low-mass X-ray binaries \citep{1988ApJ...335..755K,2002ApJ...565.1107P}.

\section{Method} \label{sec2}

We use the stellar evolution code Modules for Experiments in Stellar Astrophysics (\texttt{MESA},version 10398, \cite{2011ApJS..192....3P,2013ApJS..208....4P,2015ApJS..220...15P,2018ApJS..234...34P,2019ApJS..243...10P}) to calculate the binary evolution. Our binary evolution models begin with a NS and a zero-age main-sequence star in a circular and synchronized orbit. The initial mass of the NS is set to the canonical value of $1.4~M_\odot$, and the metallicity of the companion star is $Z=0.02$.

The mass transfer rate of Roche-lobe overflow is calculated by Kolb scheme \citep{1990A&A...236..385K}. It is generally accepted that the transferred mass cannot be fully accreted by NS, and here we adopt the isotropic re-emission model \citep{2023pbse.book.....T} to compute the mass-loss from the binary system. {The fraction of mass lost from the vicinity of NS $\beta=0.7$ is assumed.} Thus, the accretion rate of NS is $\dot{M}_{\rm NS}=(1-\beta)|\dot{M_2}|$, where $\dot{M}_2$ is the mass lost rate of companion star. $\dot{M}_{\rm NS}$ is still limited by Eddington limit.

The binary orbit can change due to angular momentum loss. We consider three mechanisms for this loss: gravitational-wave radiation, magnetic braking and mass loss. The orbital angular momentum loss due to gravitational-wave radiation is calculated by \citep{1971ctf..book.....L}
\begin{equation}
        {\dot{J}}_{\mathrm{GW}}=-{\frac{32G^{7/2}M_{\mathrm{NS}}^{2}M_{\mathrm{2}}^{2}(M_{\mathrm{NS}}+M_{\mathrm{2}})^{1/2}}{5c^{5}a^{7/2}}},
\end{equation}
where $G$ is the gravitational constant, $c$ is the speed of light in vacuum and $a$ is the semi-major axis of the orbit. The orbital angular momentum loss due to magnetic braking is calculated using standard magnetic braking prescription \citep{1983ApJ...275..713R}
\begin{equation}
    \dot{J}_{\mathrm{MB}}=-6.82\times10^{34}\left(\frac{M_2}{M_\odot}\right)\left(\frac{R_2}{R_\odot}\right)^{\gamma}\left(\frac{1~\rm day}{P_{\mathrm{orb}}}\right)^3\mathrm{dyn~cm},
\end{equation}
where $M_2$ and $R_2$ is the mass and radius of the companion star, respectively. $\gamma$ is the magnetic braking index ( we adopot $\gamma=4$ in this paper), and $\Omega$ is the spin angular velocity of the companion star. Mass loss from the binary system carries away specific orbital angular momentum. The angular momentum loss due to mass loss is calculated by
\begin{equation}
        \dot{J}_{\text{ML}}=\beta|\dot{M}_{\text{2}}|\bigg(\frac{M_{\text{NS}}}{M_{\text{NS}}+M_{\text{2}}}\bigg)^2\frac{2\pi a^2}{P_{\text{orb}}},
\end{equation}
where $P_{\text{orb}}$ is the orbital period. 

The irradiation effect on the companion star is considered in this paper and is numerically implemented by injecting extra energy into the envelope of the companion star. The accretion luminosity is given by:
\begin{equation}
        L_{\rm X}=\frac{GM_{\rm NS}\dot{M}_{\rm NS}}{R_{\rm NS}},
\end{equation}
where $R_{\rm NS}$ is the radius of the NS. Irradiation luminosity $L_{\rm irr}$ is limited by binary geometry and we follow \cite{1994ApJ...434..283H} to calculate $L_{\rm irr}$:
\begin{equation}
 \\
L_{\rm irr} =\left\{\begin{array}{l l}{{\eta L_{\rm X}\left(\frac{R_{2}}{2a}\right)^{2}}}&{{\dot{M}_{2}<\dot{M}_{\rm Edd}}}\\ {{\eta L_{\rm X}\left(\frac{R_{2}}{2a}\right)^{2}\exp\left(1-\frac{|\dot{M}_{2}|}{\dot{M}_{\rm Edd}}\right)}}&{{\dot{M}_{2}\geq\dot{M}_{\rm Edd}}}\end{array}\right., 
\end{equation}
where $\eta$ is the irradiation efficiency {, $\dot{M}_{\rm Edd}$ is the Eddington accretion limit of NS}. The extra energy is deposited in the outer layers of the companion star, and the amount of extra energy in each layer decreases exponentially from the outside to the inside of the star as $e^{-\tau}$, where $\tau$ is the optical depth at a given radius \citep{2017ApJ...847...62L}.

The spin evolution of NS is also considered and self-consistently calculated using \texttt{MESA}. Currently, the spin evolution of the NS can be described by three characteristic radii: magnetospheric radius $r_{\rm mag}$, co-rotation radius $r_{\rm co}$ and light cylinder radius $r_{\rm lc}$. {When $r_{\rm mag} > r_{\rm lc}$, the transferred matter cannot couple with the magnetosphere, causing the system to manifest as a radio binary pulsar. In the regime where $r_{\rm co} < r_{\rm mag} < r_{\rm lc}$ (commonly termed the propeller regime), while matter couples to the magnetosphere, the centrifugal force dominates over gravitational forces, resulting in matter ejection. This propeller interaction exerts a braking torque on the NS, inducing significant spin-down. Finally, when $r_{\rm mag} < r_{\rm co}$, NS can efficiently accretes material and undergo spin-up. For more details on the spin evolution mechanism, please refer to \citet{2002apa..book.....F} and \citet{2023pbse.book.....T}.} We calculate $r_{\rm mag}$, $r_{\rm co}$ and $r_{\rm lc}$ by following \citep{2002apa..book.....F}:
\begin{equation}
    r_{\rm mag}=\xi\left(\frac{\mu^{4}}{2GM_{\rm NS}\dot{M}_{\rm NS}^{2}}\right)^{1/7},
\end{equation}
\begin{equation}
    r_{\rm co} = \left(\frac{GM_{\rm NS}P_{\rm spin}^2}{4\pi^2} \right)^{\frac{1}{3}},
\end{equation}
\begin{equation}
    r_{\rm lc} = \frac{cP_{\rm spin}}{2\pi},
\end{equation}
where $\xi$ is a dimensionless constant depending on the details of disk-magnetosphere interactions. We take $\xi=0.5$ in this paper. $P_{\rm spin}$ is the spin period of NS. $\mu$ is the magnetic moment of NS and field decay due to mass accretion is calculated by \cite{1989Natur.342..656S}
\begin{equation}
        \mu=\mu_0\left(1+\frac{\Delta M_{\rm NS}}{10^{-4}M_\odot}\right)^{-1},
\end{equation}
where $\mu_0$ is the initial magnetic moment, {we set $\mu_0=10^{45}~\rm G~cm^4$. The value of $\mu_0$ corresponds to the surface magnetic field of NS, $B_0\propto\mu_0/R_{\rm NS}^3\sim10^{12}~\rm G$, which is a characteristic value for normal isolated pulsars \citep{2005AJ....129.1993M}.} $\Delta M_{\rm NS}$ is the accreted mass of NS. For other details not mentioned here, please refer to \cite{2023ApJ...956L..24L} and \cite{2024A&A...690A..88L}.

\section{Results} \label{sec3}

\begin{figure}
    \centering
    \resizebox{\hsize}{!}{\includegraphics{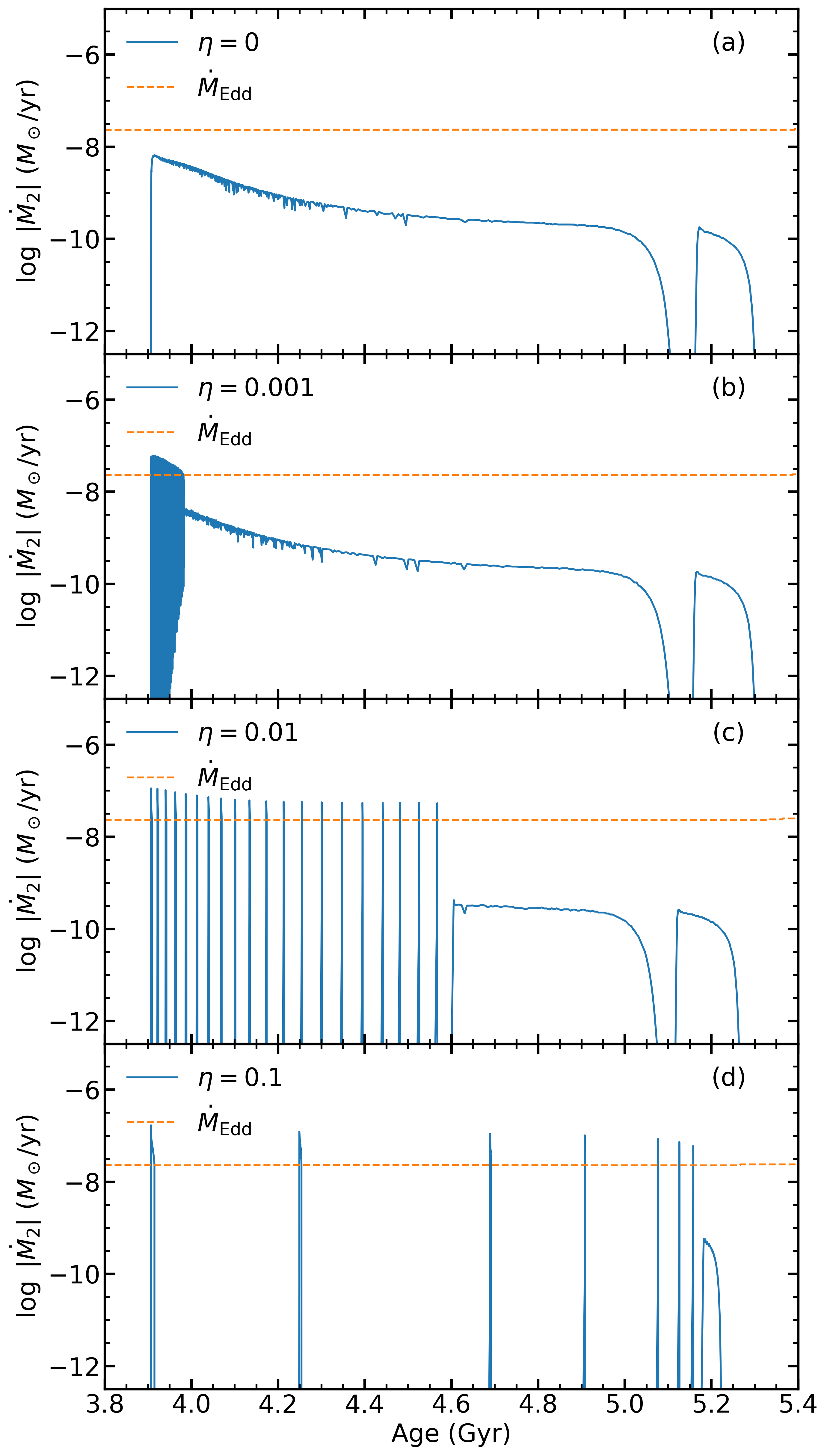}}
    \caption{Evolution of mass transfer rate as a function of time. The initial parameter for shown results is $(M_2,~\log P_{\rm orb}/{\rm days})=(1.3~M_\odot,~0.5)$. Different panels show for different irradiation efficiencies. The dashed line indicate the Eddington limit. \\
    Alt text: Mass transfer rate evolution under different irradiation efficiencies.}
    \label{fig1}
\end{figure}

We calculated a model grid for various initial orbital periods and companion masses. Here, we only show the results for $(M_2,~\log P_{\rm orb}/{\rm days})=(1.3~M_\odot,~0.5)$, as the initial parameter space for producing the observed properties of the two huntsman pulsars is quite large, as also noted by \cite{2025arXiv250105509S}. Some additional calculation results are provided in Appendix~\ref{sec_app}.

Figure~\ref{fig1} shows the evolution of the mass transfer rate. Different panels correspond to different irradiation efficiencies. In panel (a), the mass transfer is suspended for a while as the companion star enters the red bump phase at $\sim 5.1$ Gyr. In panels (b)-(d), due to the irradiation effect, mass transfer cycles emerge. As the radiation efficiency increases, the interval between successive mass transfer cycles becomes larger. The companion star would loss more mass during a single mass transfer cycle under stronger irradiation, therefore making it takes a longer time to re-fill its Roche-lobe (see also \cite{2024A&A...690A..88L}). This also means that a stronger irradiation would give the system a longer time scale in the radio pulsar binary state, i.e., in this case, huntsman state {(which can be observed as a huntsman pulsar; otherwise, the system would appear as an X-ray binary)}.

\begin{figure}
    \centering
    \resizebox{\hsize}{!}{\includegraphics{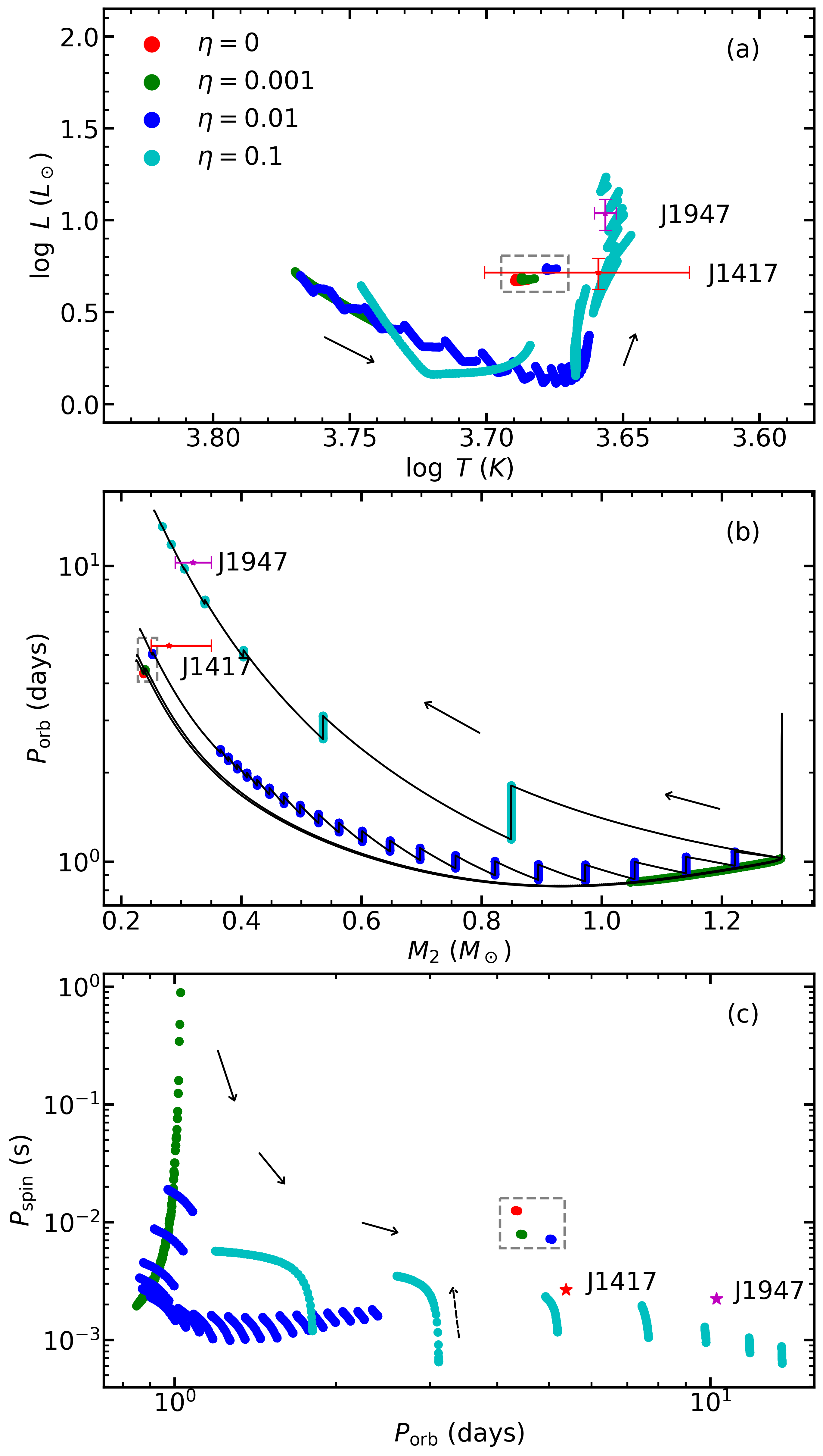}}
    \caption{Evolution results corresponding to Figure~\ref{fig1} in three parameter spaces. Panel (a): segments of evolution tracks satisfy $r_{\rm mag}>r_{\rm lc}$ in HR diagram. Panel (b): evolution tracks in companion star mass--orbital period diagram. Colored segments satisfy $r_{\rm mag}>r_{\rm lc}$. Panel (c): similar to panel (a) but in orbital period--spin period diagram (the so-called Corbet diagram). The initial magnetic moment of NS is $\mu_{\rm 0}=1.0\times10^{30}~{\rm G~cm^3}$. Different color indicate for different irradiation efficiencies. The colored dots in dashed box indicate that the stars locate at the red bump phase. {The solid arrows indicate the general evolutionary direction of the line segments, while the dashed arrow in panel (c) marks the evolutionary direction for an individual segment.} \\ 
    Alt text: Modeled evolutionary tracks in other parameter spaces to compare with observed objects.}
    \label{fig2}
\end{figure}

In Figure~\ref{fig2}, we show the evolution results in three parameter spaces. All colored dots in the three panels indicate $r_{\rm mag}>r_{\rm lc}$, meaning that the NS can manifest itself as a radio pulsar. The different colors represent different irradiation efficiencies. The panel (a) of Figure~\ref{fig2} shows our calculated results and the observed properties of the two huntsman pulsars' companion stars in the Hertzsprung–Russell (HR) diagram. Panel (a) clearly shows that the companion stars of the two huntsman pulsars may be normal red giant stars, rather than just red bump stars. For the companion star of J1417, both red giant and red bump star scenarios are possible due to its large error bars in its effective temperatures, while for J1947, a red bump companion is unlikely under the initial condition of $(M_2,~\log P_{\rm orb}/{\rm days})=(1.3~M_\odot,~0.5)$. It is just more likely to be a normal red giant star. In panel (b), we show evolutionary tracks of the system with $(M_2,~\log P_{\rm orb}/{\rm days})=(1.3~M_\odot,~0.5)$ under different irradiation efficiencies in an $M_2-P_{\rm orb}$ plane. The vertical segments marked with colored dots indicate $r_{\rm mag}>r_{\rm lc}$, which are caused by {irradiation-induced detachment phases (i.e., intervals between mass transfer cycles)} or the companion star entering the red bump phase (dashed box) (see also \cite{2024A&A...690A..88L}). The properties of J1947 can be well explained by the results for $\eta=0.1$. However, for J1417, a value of $\eta$ in the range of $0.01-0.1$ may be better to explain its properties under the initial condition of $(M_2,~\log P_{\rm orb}/{\rm days})=(1.3~M_\odot,~0.5)$. Panel (c) shows a $P_{\rm orb}-P_{\rm spin}$ diagram. The results corresponding to red bump phase (dashed box) are difficult to reach the spin levels of two huntsman pulsars. A larger $\eta$ may help the NS achieve a lower spin period. The results for $\eta=0.1$ are sufficient to explain the spin properties of the two huntsman pulsars. However, for J1417, $\eta=0.1$ can not explain the companion mass at the same time from panel (b). Therefore, a value of $\eta$ in the range of $0.01-0.1$ is still a better choice for J1417. It should be noted that our calculated spin periods should be regarded as lower limits, as additional spin-down mechanisms, such as gravitational radiation, are not considered here. Overall, our model can well explain the observed properties of the two Huntsman pulsar binaries.
\section{Discussion and Conclusion} \label{sec4}

As mentioned earlier, the initial parameter space for producing the observed properties of the two huntsman pulsars is quite large. In Appendix~\ref{sec_app}, we provide some additional calculation results. In our models, one can vary initial parameters, such as the irradiation efficiency, companion star mass, and orbital period, to easily reproduce the observed properties of the two huntsman pulsars. For example, as shown in Figure~\ref{app_fig2}, increasing the initial companion mass results in a smaller final orbital period, making it harder to explain J1417. Similarly, in Figure~\ref{app_fig1}, increasing the initial orbital period leads to a larger final orbital period, and neither huntsman pulsar can be explained. Which means there some fine tuning problems could exist to reproduce the huntsman pulsars in initial orbital periods as noticed in some pulsars with low mass helium WD \citep{2014A&A...571A..45I}.

We have shown that our model, which includes the irradiation effect, can explain the observed properties of the two huntsman pulsar binaries, J1417 and J1947, including the luminosity, effective temperature, and mass of the companion stars, as well as the orbital and spin periods. From the perspective of our results, it is difficult to spin up the NS to the level observed in the two huntsman pulsars without the irradiation effect. In models without irradiation, when the companion star enters the red-bump phase, the mass transfer is suspended for a while due to the shrinking of the companion star. The so-called Roche-lobe decoupling phase can significantly spin down the NS \citep{2012Sci...335..561T}. {As the donor star decouples from its Roche lobe, the mass transfer rate decreases at the same time. This causes the NS to transition from the accretion regime to the propeller regime. The associated braking torque induces significant spin-down prior to the system's evolution into the radio pulsar phase.} Such phenomenon also related to the initial magnetic field strength of NS, which still needs detailed investigation \citep{2024MNRAS.535..344K}. Therefore, using the observed spin period of J1947 alone to estimate the accreted mass of the NS is not very accurate\footnote{See also Figure~4 in \cite{2021ApJ...922..158L}}. A detailed calculation is preferable. The spin evolution of an accreting NS is complex. If J1417 and J1947 are indeed products of irradiation-induced mass transfer cycles, then the spin derivatives of these two huntsman pulsars could be larger than those of pulsars with the same spin period, where the strength of initial magnetic field of NS could play a key role \citep{2023ApJ...956L..24L}.

In addition to the irradiation effect, evaporation also plays an important role in compact binary evolution. {Evaporation describes the mass stripping of the donor star's caused by the radio pulsar's high-energy particle wind. This energetic process enables additional mass loss even in detached state.} Previous works have attempted to use this mechanism to explain the origin of J1417, as it exhibits some characteristics of spider pulsars \citep{2019MNRAS.483.4495D,2021MNRAS.500.1592G,2023MNRAS.525.2708C}. However, the spin evolution of the NS was not included in previous studies. Similarly, we should incorporate evaporation into our models in future work.

Another interesting aspect of our model is that the companion star can be a normal red giant rather than solely a red bump star, providing an additional way to verify our irradiation model. {For the donor star, a potential approach may be examining archival data to check long-term luminosity trends. Sustained brightness decrease may support the red bump scenario, whereas stable increasing luminosity may support the irradiation scenario.} Even if the companion stars of J1417 and J1947 are later confirmed to be red bump stars, our results suggest that at least a weak irradiation effect should be considered in binary evolution. This effect would cause the companion to expand slightly, increase the mass transfer rate, and weaken the impact of the Roche-lobe decoupling phase, ultimately facilitating the acceleration of the NS to a shorter spin period.

In any case, the discoveries of J1417 and J1947 are important for both single and binary star evolution. Finding more huntsman pulsars can significantly constrain binary evolution theory and enhance our understanding of NS spin evolution.

\begin{ack}
We sincerely thank the anonymous referee for their constructive feedback, which greatly improved the manuscript. The authors also acknowledge the ``PHOENIX Supercomputing Platform'' jointly operated by the Binary Population Synthesis Group and the Stellar Astrophysics Group at Yunnan Observatories, Chinese Academy of Sciences.
\end{ack}

\section*{Funding}
This work is supported by the National Natural Science Foundation of China (Nos. 12288102 and 12333008), National Key R\&D Program of China (No. 2021YFA1600403) and the Strategic Priority Research Program of the Chinese Academy of Sciences (grant Nos. XDB1160303, XDB1160000). X.M. acknowledges support from Yunnan Fundamental Research Projects (Nos. 202401BC070007 and 202201BC070003), International Centre of Supernovae, Yunnan Key Laboratory (No. 202302AN360001), the Yunnan Revitalization Talent Support Program-Science \& Technology Champion Project (NO. 202305AB350003), and the science research grants from the China Manned Space Program with grant no. CMS-CSST-2025-A13.

\appendix 
\section{More evolution results} \label{sec_app}

\begin{figure}
    \centering
    \resizebox{\hsize}{!}{\includegraphics{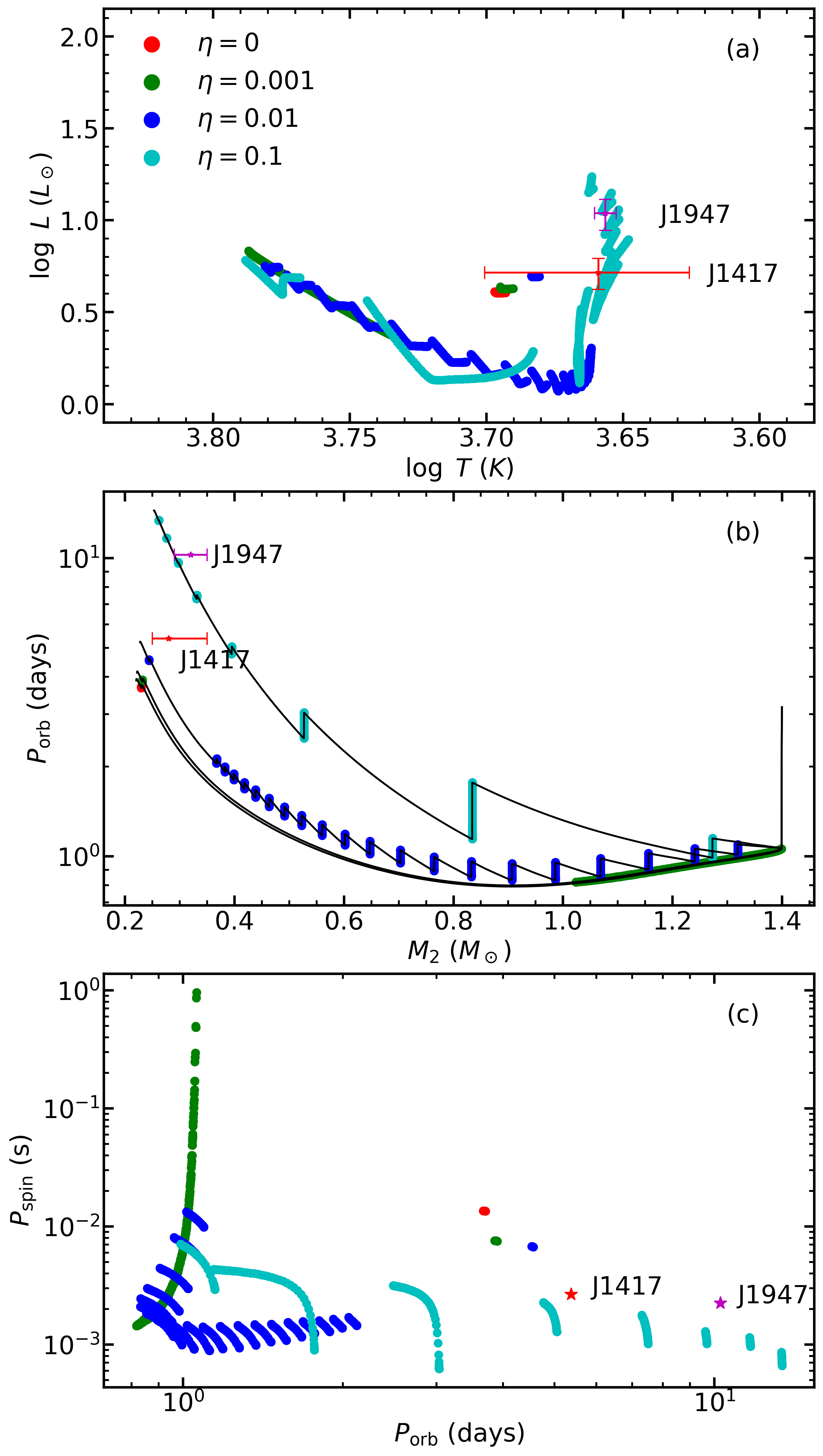}}
    \caption{Same as Figure~\ref{fig2}, but $(M_2,~\log P_{\rm orb}/{\rm days})=(1.4~M_\odot,~0.5)$. \\
    Alt text: Modeled evolutionary tracks with a different initial parameters compare to Figure~\ref{fig2}.}
    \label{app_fig2}
\end{figure}

\begin{figure}
    \centering
    \resizebox{\hsize}{!}{\includegraphics{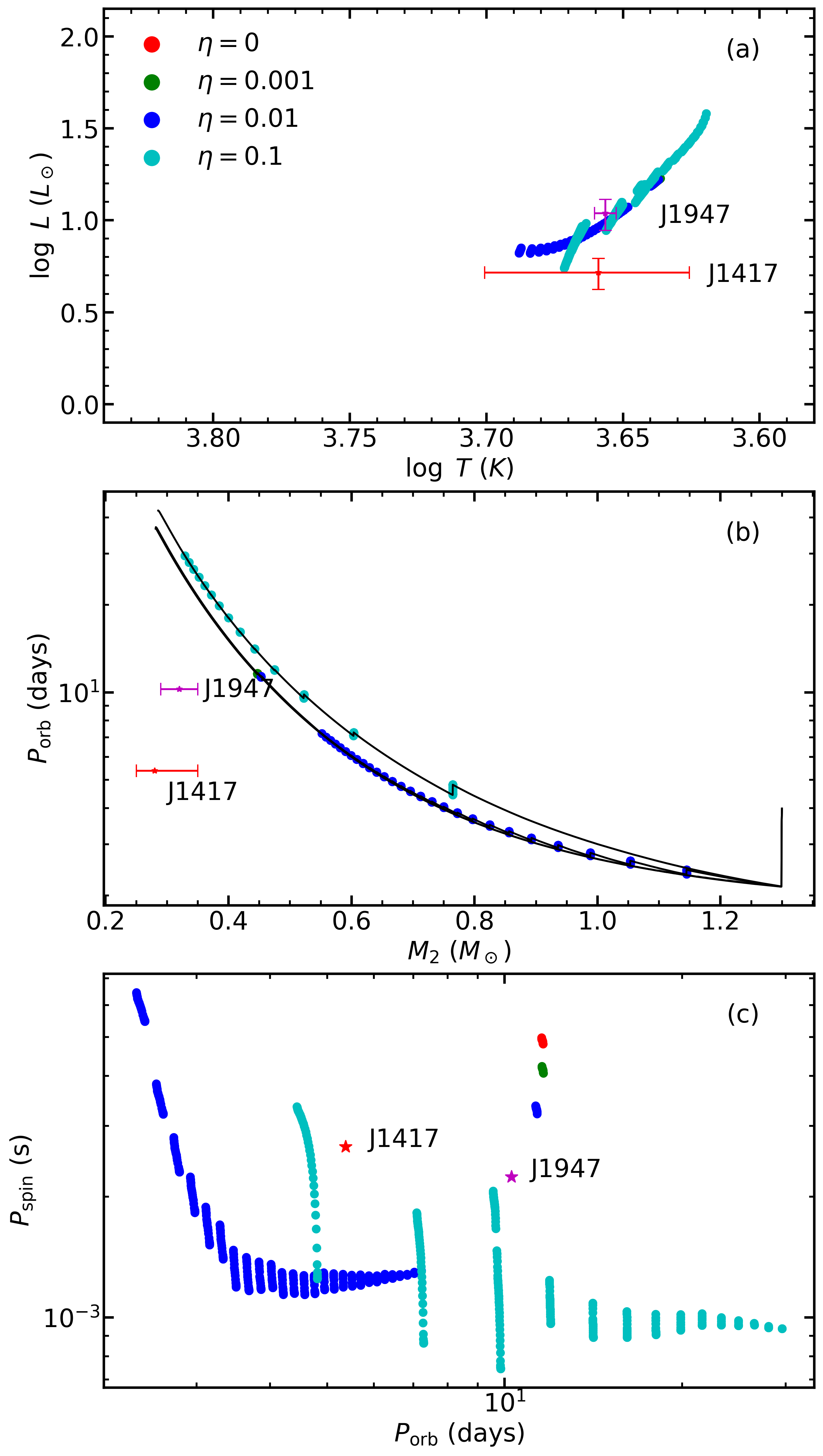}}
    \caption{Same as Figure~\ref{fig2}, but $(M_2,~\log P_{\rm orb}/{\rm days})=(1.3~M_\odot,~0.6)$. \\ 
    Alt text: Modeled evolutionary tracks with another different initial parameters compare to Figure~\ref{fig2}.}
    \label{app_fig1}
\end{figure}

In this section, we show some additional evolution results. We change the initial mass of companion and orbital period compare to Figure~\ref{fig2}. The initial parameter $(M_2,~\log P_{\rm orb}/{\rm days})$ of Figure~\ref{app_fig2} is $(1.4~M_\odot,~0.5)$, and $(1.3~M_\odot,~0.6)$ for Figure~\ref{app_fig1}. See Section~\ref{sec4} for more discussion.

\bibliography{main}{}
\bibliographystyle{aasjournal}

\end{document}